\documentclass[structabstract]{aa}  
\usepackage{graphicx}
\usepackage{txfonts}
\usepackage[]{natbib}

\usepackage{multirow}

\titlerunning{Characterization of horizontal flows around solar pores}
\authorrunning{S. Vargas Dom\'inguez et al.}

\def\farcs{\hbox{$.\!\!^{\prime\prime}$}}
\def\arcsec{\hbox{$^{\prime\prime}$}}

\begin{document}
   \title{Characterization of horizontal flows around solar pores from high-resolution time series of images}

   \author{S. Vargas Dom\'inguez,
          \inst{1}
          A. de Vicente \inst{2},   J.A. Bonet \inst{2}
          \and
       V. Mart\'inez Pillet\inst{2}
          }

   \institute{Mullard Space Science Laboratory, University College London,
              Holmbury, St Mary, Dorking, RH5 6NT, UK\\
              \email{svd@mssl.ucl.ac.uk}
         \and
             Instituto de Astrof\' isica de Canarias, 38205  La Laguna, Tenerife, Spain\\
                      }

   \date{Received September 8, 2009; accepted month day, 2009}

  
  \abstract
   {Though there is increasing evidence linking the moat flow and the Evershed flow
along the penumbral filaments, there is not a clear consensus regarding the
existence of a moat flow around umbral cores and pores, and the debate is still
open. Solar pores appear to be a suitable scenario to test the moat-penumbra
relation as evidencing the direct interaction between the umbra and the
convective plasma in the surrounding photosphere, without any intermediate
structure in between. }   
   {The present work studies solar pores based on high
resolution ground-based and satellite observations.}
   {Local correlation tracking
techniques have been applied to different-duration time series to analyze the
horizontal flows around several solar pores.}
   {Our results establish that the
flows calculated from different solar pore observations are coherent among each
other and show the determinant and overall influence of exploding events in the
granulation around the pores. We do not find any sign of moat-like flows
surrounding solar pores but a clearly defined region of inflows surrounding
them.}
   {The connection between moat flows and flows associated to penumbral
filaments is hereby reinforced by this work.}


   \maketitle
%

\section{Introduction}

The solar photosphere displays a wide variety of magnetic features at different
spatial scales being sunspots and pores the more conspicuous ones. The
morphology and evolution of sunspots and pores have been extensively studied
and both are thought to be ruled by the mutual effect of magnetic field that
inhibits convection (though not completely) and plasma motions. Nevertheless,
there is not clear consensus for a model explaining the transition from pores
into sunspots (i.e.\ development of penumbra, see the monograph
by \cite{thomasweiss08}), the fade of the Evershed flow
once leaving the penumbra and entering the region dominated by the large
outflows (the so-called moat flows), the flow patterns surrounding pores
which are, alternatively, dominated by downflows surrounding them \citep{giordano08}
or moat-like flows \citep{zuccarello09} and, also, the effect of the penumbra in the
granular pattern surrounding sunspots.\\

By using local correlation tracking techniques \cite{vargas2007} found a direct
correlation between the presence of penumbrae and the appeareance of moat flows
in a complex $\delta$-configuration active region. A more extensive sample
taking into account different penumbral configurations was analyzed by
\cite{vargas2008}, establishing a systematic moat-penumbra relation in all the
sunspots under study. According to these studies,  no moat flow was detected in
the granulation next to umbral boundaries lacking penumbrae. Moat flows were
always detected as a prolongation of the penumbral filaments once crossing the
penumbral boundary. \\

Though there is increasing evidence linking the moat flow and the Evershed flow
along the penumbral filaments \cite[e.g.][]{sainz2005,cabrera2006}, the debate
regarding the existence of a moat flow around umbral boundaries without
penumbra and individual pores is still ongoing. In a recent work,
\cite{deng2007}, found that the dividing line between radial inward and outward
proper motions in the inner and outer penumbra, respectively, survived the
decay phase, suggesting that the moat flow is still detectable after the
penumbra disappeared. However, previous works \citep{sobotka1999, roudier2002,
hirzberger2003} have measured horizontal proper motions in and around pores and
have observed some penetrating flows at the umbral boundaries and a ringlike
arrangement of positive divergences (\emph{rosettas}) around the pores which is
related to a continuous activity of exploding granules in the granulation
around them. \cite{roudier2002} identified a very clear inflow around pores
which corresponds to the penetration of small granules and granular fragments
from the photosphere into the pores, pushed by granular motions originated in
the divergence centres around them. These authors conclude that the motions at
the periphery of the pore are substantially and continuously influenced by the
external plasma flows deposited by the exploding granules. It is important to
note in this context that the annular area surrounding pores and filled with 
exploding granules generates an outward directed flow annulus that can give
the impression of a persistent outflow.  \\

Converging flows around pores have also been observationally reported previously by \cite{wang1992} as well as downflows at their periphery \citep{sankara2003}. More recently, 3D magnetohydrodynamic simulations found horizontal flows  towards the pores that contribute to mantain their magnetic structure together  \citep{cameron2007}.\\

\begin{table*}
\centering
\caption{Characteristics of the time series of solar pores observed from ground-based and satellite facilities.}
\begin{tabular}{cccccccc}\\
Telescope & Date 2007  & Series & Time, UT& Duration \footnotesize[min] & N. images & Cadence \footnotesize[sec] & FOV \footnotesize[$\arcsec$]\\\hline\hline\\
\multirow{2}{*}{\emph{SST}} & \multirow{2}{*}{30 Sep} & 1 & 08:43-09:31 & 48 & 286 & 10 & 64.3$\times$65.0  \\
& & 2 & 09:36-09:56 & 20 & 118 & 10 &  64.3$\times$65.0 \\\hline\\
\multirow{3}{*}{\emph{Hinode}} & \multirow{2}{*}{1 Jun} & 1 & 21:35-21:55 & 20 & 40 & 30 & 27.9$\times$55.7 \\
& & 2 & 22:26-23:33 & 67 & 134 & 30 &  27.9$\times$55.7 \\\cline{2-8}\\
& 30 Sep& 1 & 00:14-17:59 &  960 & 1030 & 60 & 55.7$\times$111.4  \\\hline
\end{tabular}
\label{poroseries}
\end{table*}

\begin{figure}
\vspace{-1.5cm}
\includegraphics[width=1.\linewidth]{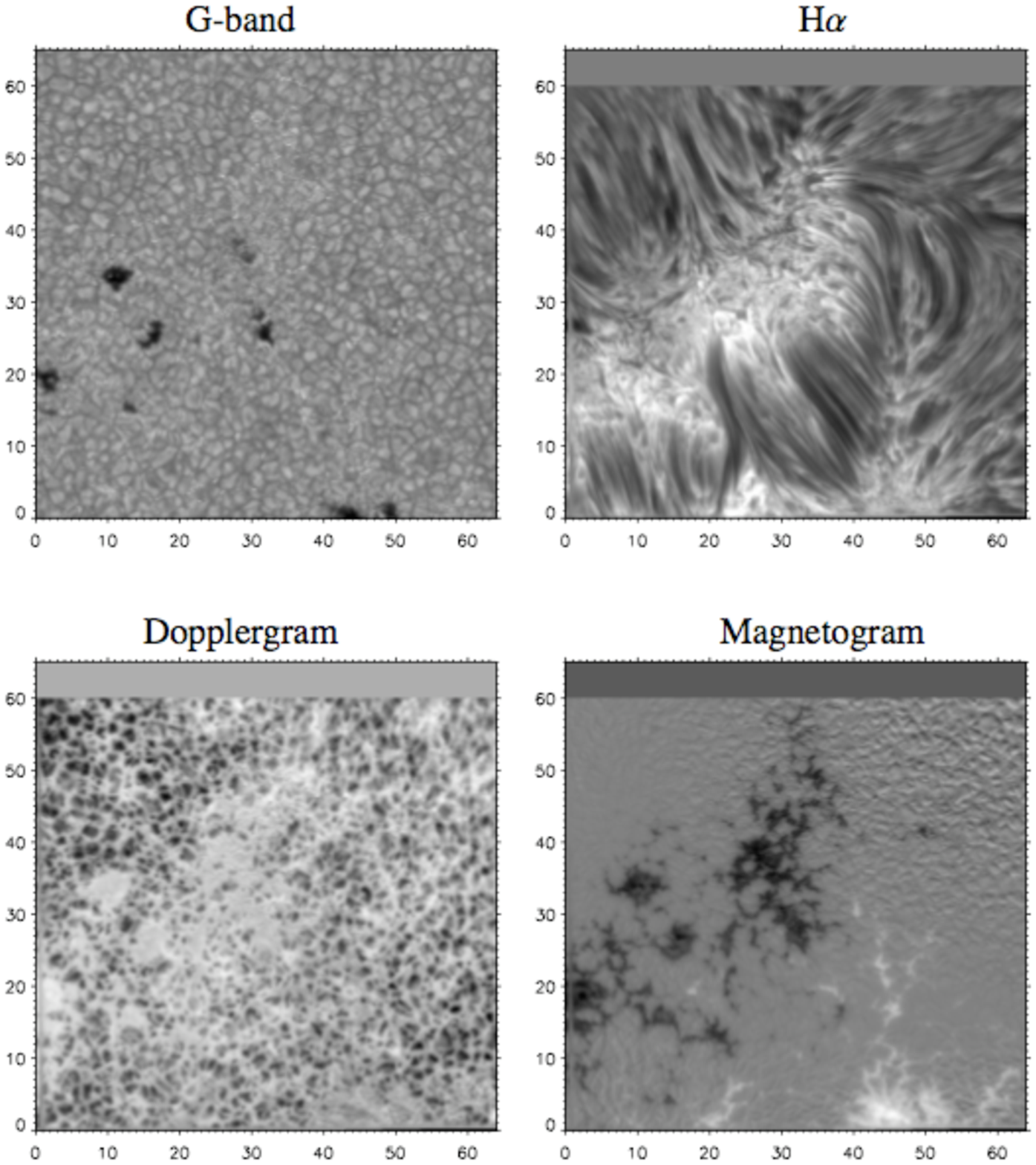} 
\vspace{-2cm}
\caption{SST co-temporal and co-spatial set of images from the emerging flux region
on 30 September 2007. Dark regions in the Dopplergram denote upflows and the map ranges from approx. -1.2 (\emph{white}) to 1.6  (\emph{black}) km s$^{-1}$.
The LOS Magnetogram shows negative/positive (\emph{black/white}) polarities with field strengths up to$~\sim$2300 G. The coordinates are expressed in arc sec.}
\label{images_red}
\end{figure}

Pores are interesting to analyze since, as they do not display penumbrae
\citep{keil1999}, what we actually observe is the direct interaction between
the umbra (with a strong vertical magnetic field that inhibits convection
inside it) and the convective plasma in the surrounding photosphere, without
any intermediate structure in between. Many observed features such as bright
granules moving in the border of a pore \citep{sobotka1999} show the complex
exchange taking place between the pore and its surrounding granulation.\\

Our main interest is the characterization of the horizontal flows around a variety of solar pores on the
basis of high-resolution time series of images. Observations from ground-based
and space telescopes are analyzed by means of the local correlation tracking
technique. In Sect.~2, the paper concentrates on the description of the images
acquisition and data processing separately for ground-based and satellite data.
The analysis of the data and the presentation of results are treated in Sect.~3. A
general summary and final discussion are presented in Sect.~4.\\

\section{Observations and data processing}
\label{sec:obs}

A significant part of our data were acquired during a long observing campaign (24 days)
carried out  in September-October 2007 with the cooperation of several European
and Japanese institutions and joint observations from several solar telescopes
of the Canary Islands Observatories: SST and DOT in La Palma and VTT and THEMIS
in Tenerife. Moreover, and for the very first time, coordinated observations
with the space solar telescope \emph{Hinode}  \citep{kosugi2007} were performed
in the framework of the \emph{Hinode Operation Program 14}. In the next
sections, we will detail the ground-based and satellite observations supporting
the present work as well as the specific data processing applied in each
case.\\

\begin{figure}
\hspace{-2cm}\includegraphics[width=1.4\linewidth]{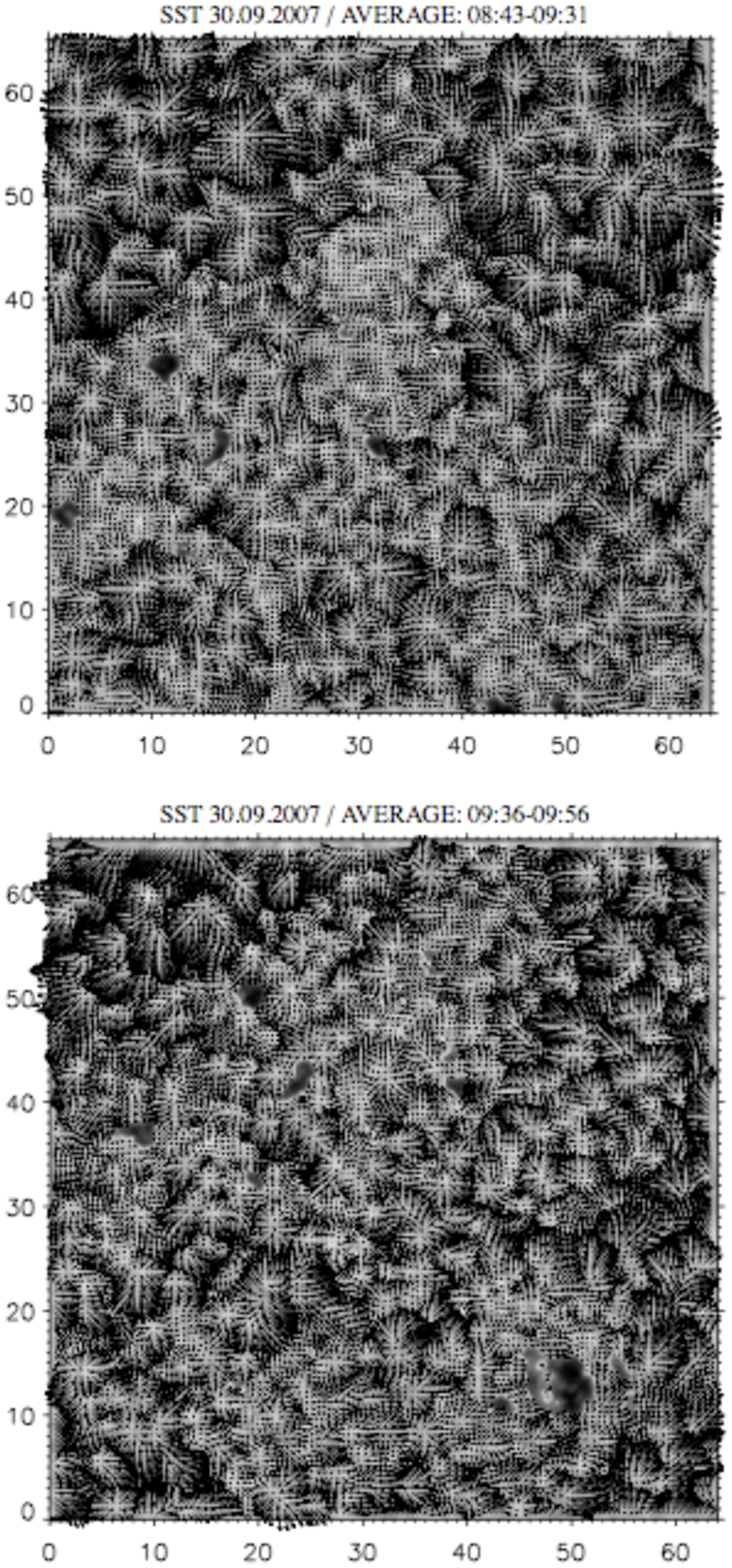} 
\caption{Map of horizontal velocities (FWHM 1$\farcs$0) from the restored time
series taken at the SST on September 30, 2007. The velocities are averaged over
48 minutes (\emph{upper panel}) and 20 minutes (\emph{lower panel}). Arrows show the direction 
of horizontal velocities and their length the correspondent absolute value in all flow maps hereafter. The \emph{white contours} outline the border of solar pores. The length of the
black bar at coordinates (0,0) corresponds to 1.6 km s$^{-1}$. The coordinates
are expressed in arc sec. The background in every case represents the average
image of the corresponding G-band series.}
\label{porosflowmapSST}
\end{figure}

\subsection{Ground-based SST data}
\label{S:obsSST}

The data from the Swedish 1-m Solar Telescope \citep[SST,][]{SSTa,SSTb}
analyzed in the present work were recorded during a particular observing run on
30 September 2007 and correspond to the active region NOAA 10971. The main
target was a region close to the solar disc center ($\mu$=0.98) with some pores
of different sizes embedded in a plage region that exhibits an intense magnetic
activity.\\

A dichroic beam-splitter in the optical setup divided the light beam into two
channels: blue and red. Images in G-band ($\lambda 4305.6$~\AA) were acquired 
in the blue beam at a rate of ~13 frames s$^{-1}$. We used 12 bit detectors of 
$2048 \times 2048$ square pixels which in
combination with an image-scale of 0.034 arcsec/pix rendered an effective
field-of-view (hereafter FOV) of  $69 \times 69$ square arcsec. After dark
current subtraction and flatfielding, G-band images were corrected for atmospheric 
and instrumental degradation by employing the so-called 
\emph{Multi-Frame Blind-Deconvolution} restoration
technique \citep[MFBD,][]{lofdahl1996,lofdahl2002MFBD}. For each wavelength,
the image sequence was grouped in sets of about 80 consecutive frames acquired
within time intervals of 10 seconds each. Every set yielded one restored image.
Next steps in the data post-processing were: compensation for diurnal field rotation, rigid alignment
of the images, correction for distortion and finally subsonic filtering. For
more details see \cite{vargasthesis}.\\

The red beam fed the filter  \emph{Solar Optical Universal Polarimeter}
\citep[SOUP,][]{title1986} to obtain H$\alpha$ ($\lambda 6562.8$~\AA) and
narrow-band images at $6302$~\AA. Single images were taken at a rate of 35
frames s$^{-1}$ with CCDs of $1024 \times 1024$ square pixels and an
image-scale of 0.065 arcsec/pix. A beam splitter in front of the SOUP deflected
a fraction of light to obtain simultaneous broad-band Phase Diversity
image-pairs in the continuum near $\lambda6302$ \AA.  These images made up an
additional \emph{object} for the \emph{Multi-Object Multi-Frame
Blind-Deconvolution} \citep[MOMFBD,][]{MOMFBD} algorithm to jointly restore
both, the broad-band and the narrow-band images. From the restored narrow-band
images, we computed longitudinal magnetograms and dopplergrams.\\

Due to periods of bad seeing in which the quality did not reach the desired top
level, some of the images were discarded and we kept only the best and longest
consecutive sequence of images. The very final product were two G-band time series 
with the characteristics listed in Table~\ref{poroseries}. 
The time gap of about 5 min between the two consecutive
series for each wavelength resulted from a telescope tracking interruption.
That is also the reason why the FOV is slightly different in the time series.
Most images forming the time series show details near the diffraction limit of
the telescope.\\

The procedure followed for image restoration is extensively detailed in
\cite{vargasthesis}. Figure~\ref{images_red} shows one of the co-temporal sets
of images of the emerging active region after restorations. The upper left panel shows the
FOV covered by the observations displaying the pores under analysis and the other panels useful information to characterize the region as derived from the red channel simultaneous observations.\\

\begin{figure}
\hspace{-1cm}\includegraphics[width=1.3\linewidth]{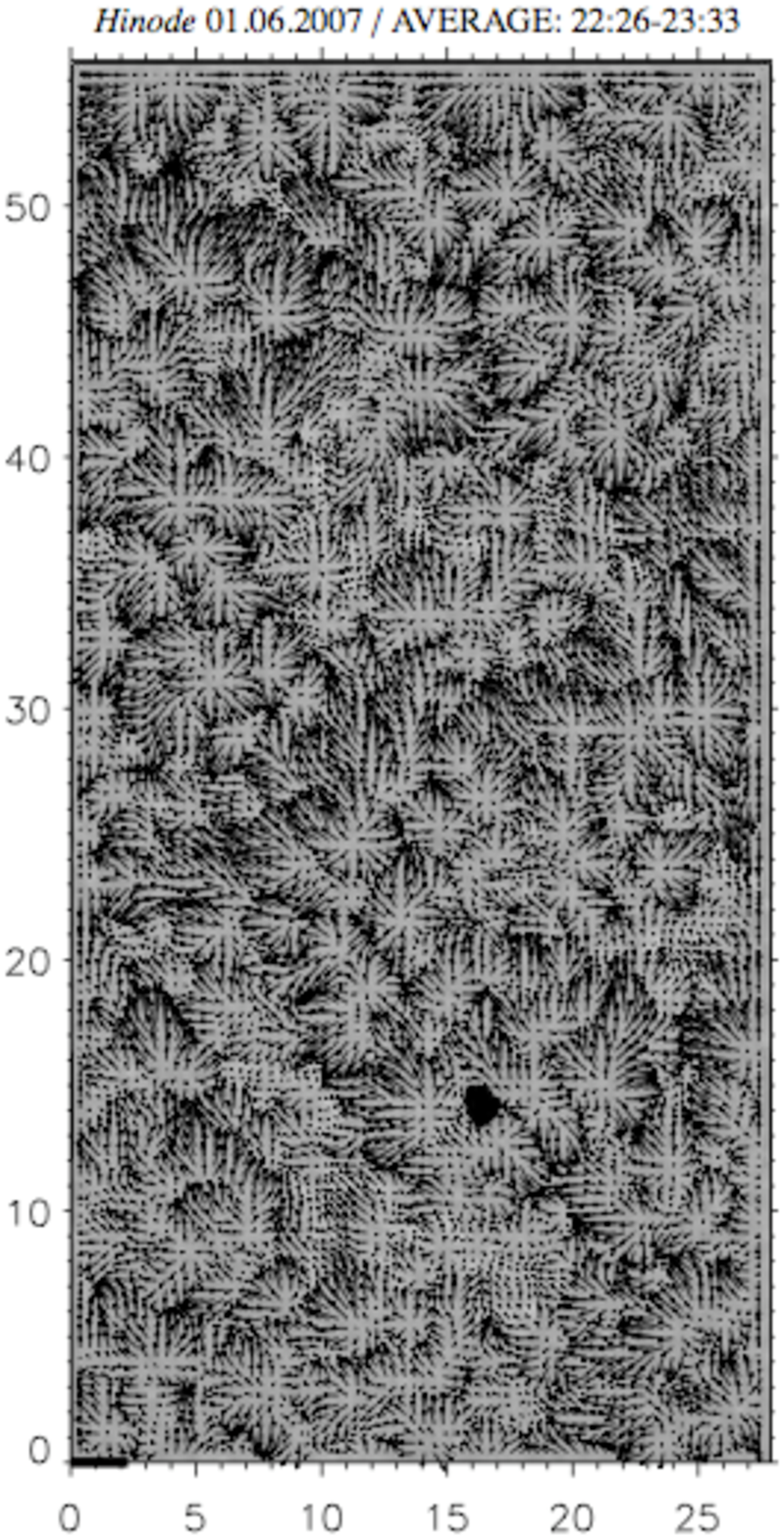} 
\vspace{-0.5cm}
\caption{Maps of horizontal velocities (FWHM 1$\farcs$0, 67 minutes average)
for the processed time series taken by \emph{Hinode} on June 1, 2007. The
length of the black bar at coordinates (0,0) corresponds to 2.3 km s$^{-1}$.
The coordinates are expressed in arc sec. The background represents the average
image of the G-band series.}
\label{porosflowmapHinode}
\end{figure}

\subsection{Hinode space satellite data}
\label{S:obsHinode}

After its launch on 22 September 2006, \emph{Hinode} {has become an extremely successful
observatory for} 
Solar Physics research. It has managed to observe many high-detailed
solar features by avoiding the blurring and distortion effects produced by the
Earth's atmosphere. The public archive of \emph{Hinode}\footnote{See the
website: \emph{http://solar-b.nao.ac.jp/hsc\_e/darts\_e.shtml}} is an organized
data base where all \emph{Hinode} observations can be found out and easily
downloaded. \\

We were interested in \emph{Hinode} observations of solar pores taken with the
Solar Optical Telescope \citep[SOT,][]{tsuneta2008}, in order to pursue our
study of photospheric horizontal flows. The next two sections describe the
\emph{Hinode} data we analyze in this paper. These data extend the sample of
cases under study in the present work and moreover give us the possibility to
compare with the results stemming from ground-based observations.\\

\subsubsection{Data from 1st June 2007}
\label{1jun}

These data correspond to a solar portion including an isolated and round-shaped
pore observed by \emph{Hinode} on 1st June 2007. Images were acquired in G-band
with a cadence of 30 seconds in a solar portion close to the disk center
($\mu$=0.87). After subsonic filtering we obtained two time series;
Table~\ref{poroseries} (1 Jun) summarizes the parameters of both series in
detail. The treatment performed on \emph{Hinode} data does not include any
restoration process because they were not degraded by atmospheric turbulence.\\

\subsubsection{Data from 30 September 2007}
\label{30sep}

The data from \emph{Hinode} on 30 September 2007 correspond to the coordinated
observations described in Sect.~\ref{S:obsSST}. \emph{Hinode} observed in G-band
the emerging flux region NOAA 10971 from 00:14 to
17:59 UT with a few brief interruptions for calibrations.
Table~\ref{poroseries} (30 Sep) summarizes the parameters of the time series in
more detail.  A significant part of the FOV in the SST images is covered by the
\emph{Hinode} observations.\\

\emph{Hinode} data often presents misalignments due to tiny tracking
flaws and also due to temporal interruptions. We proceeded by aligning all of the
1030 images (18 hours) at a sub-pixel level. Subsonic filtering was applied to 
eliminate residual jittering in the time series.\\

\begin{figure}
\vspace{-3.5cm}
\includegraphics[width=1.\linewidth]{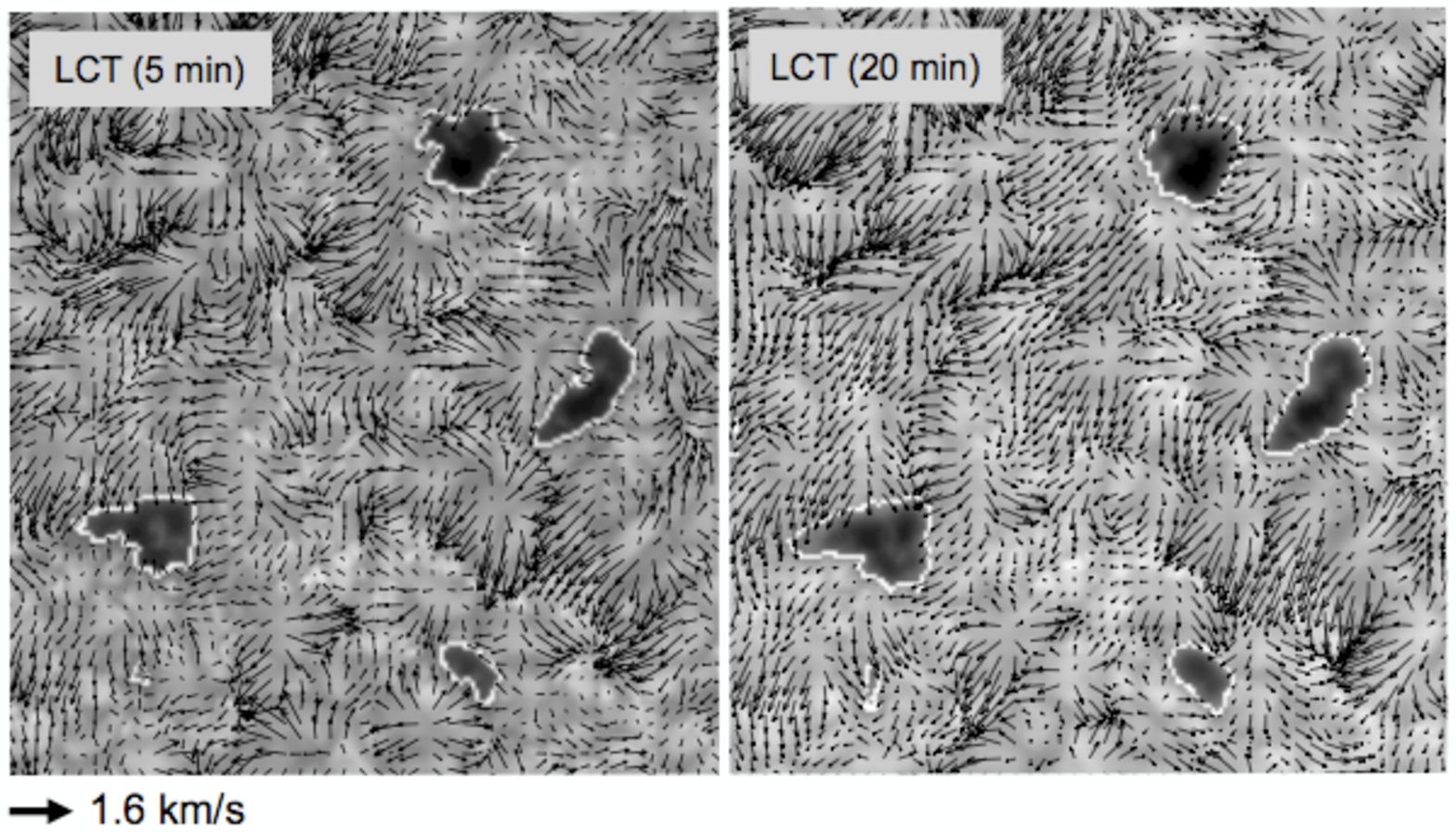}
\vspace{-4cm}
\caption{Maps of horizontal velocities (FWHM 1$\farcs$0) for the SST time series on September 30, 2007.
Velocities are derived from LCT for two different time averages as labeled. The background in every case represents the average image of the corresponding G-band series. The FOV
of every image is 25\arcsec$\times$27\arcsec.}
\label{LCTaver}
\end{figure}

\begin{figure}
\centering
\hspace{-6mm}
\hspace{-2mm}\includegraphics[width=1.\linewidth]{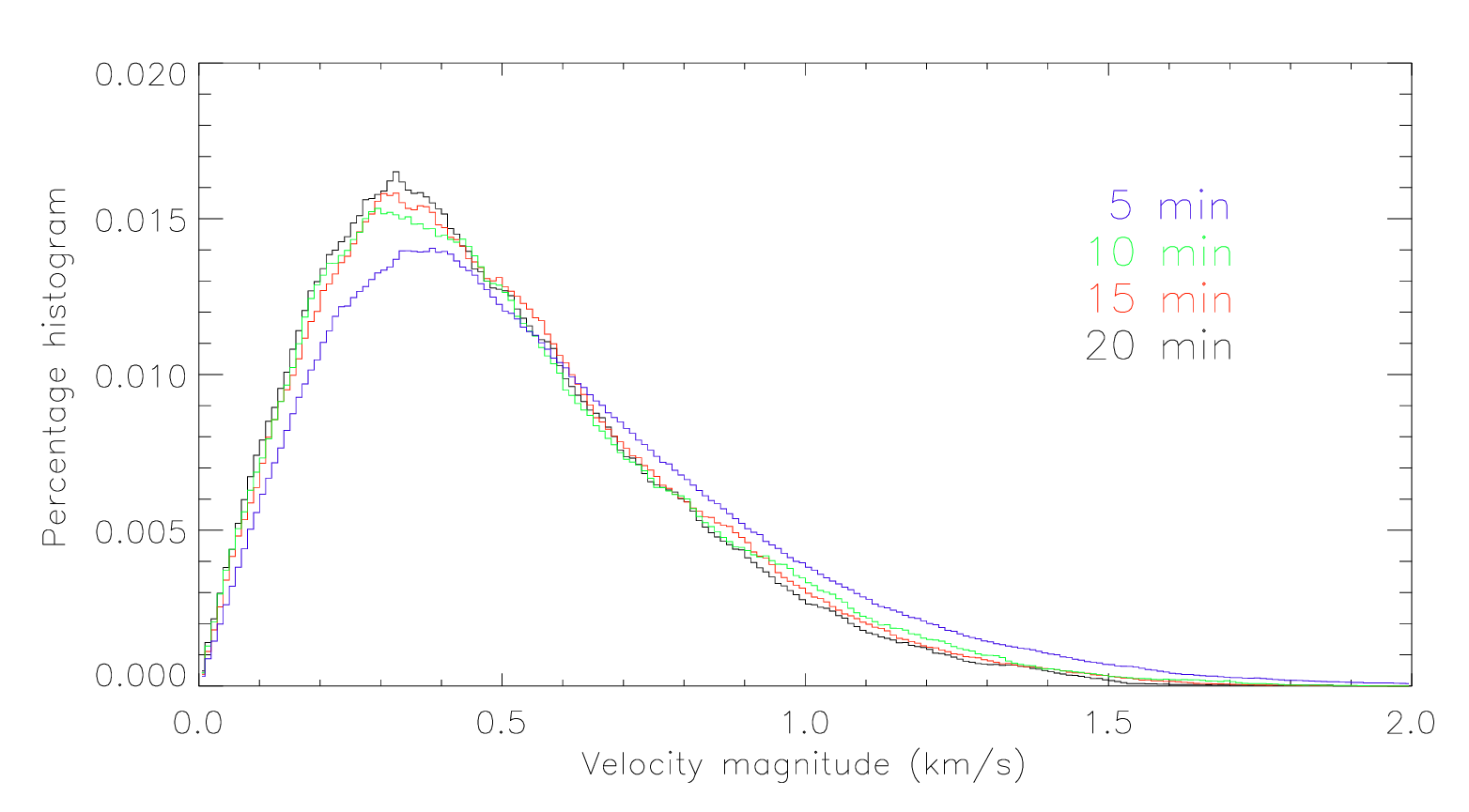}
\caption{Histogram of horizontal velocity magnitudes for different time
intervals. A local correlation tracking technique has been used to derive the
horizontal velocities (FWHM=1$\farcs$0).}
\label{histogram_averages}
\end{figure}

\section{Data analysis and results}
\label{S:res}

\subsection{General description of horizontal proper motions in the FOV}
\label{horizontalpattern}
The G-band series taken at both SST and \emph{Hinode} have been employed to
analyze the horizontal proper motions of structures in every FOV, by using the
local correlation tracking (LCT) technique \citep{november1988} as implemented
by \cite{molowny1994}. This technique works by selecting small subfields around the same
pixel in contiguous snapshots, which are correlated to find the best match-displacement. 
The procedure provides a map of displacements or proper motions  per time step (i.e. velocities),
which we average in time.  The subfields to be correlated are defined by a Gaussian window whose 
full-width at half-maximum (FWHM) is set according to the size of the structures we want to track.
In this section, we show the maps of horizontal
velocities calculated for the different time series by using a Gaussian
tracking window of FWHM 1$\farcs$0 which is roughly half of the typical
granular size.\\
 
Figure~\ref{porosflowmapSST} shows the flow maps computed from the two restored
time series from SST. The velocities were averaged over the total duration of
each series. The underlying background in the maps is the average image of the
respective series. As commented above, the FOV is slightly different in both
cases so that, for instance, the pore at coordinates [17$\arcsec$,26$\arcsec$] in the upper map of
the figure is located at [24$\arcsec$,42$\arcsec$] in the lower map.\\

The maps are dominated by flows coming from exploding events taking place all
over the FOV. As expected, the map averaging over a longer time period (48 min)
is slightly smoother than the other one (20 min), and displays lower
velocities. Nevertheless, both maps reproduce similar flow patterns all over
the FOV. The top of the FOV in Fig.~\ref{porosflowmapSST} (\emph{upper
panel}) shows very conspicuous exploding granular events which are grouped at
every upper corner of the FOV forming two large-scale structures that fit well
the supergranular one. Interestingly, there is an evident anti-correlation between the strength of the
horizontal velocity field and the mean magnetic flux density in those locations as shown in the 
co-temporal and -spatial magnetogram in Fig.~\ref{images_red}. A smaller portion of these structures can also be identified in the lower panel of Fig.~\ref{porosflowmapSST}. \\

Even though a complete description of the proper motions around the pores in the FOV will be done in Sect.\ref{velpores}, a
glance at the figure reveals no evidence of a moat-like pattern around any of
the pores. The central part of the FOV where the smaller pores are embedded,
exhibits a lower magnitude of horizontal velocities. This behavior is explained
by the intense magnetic activity in this part of the FOV as unveiled by the
corresponding magnetogram of the zone shown in Fig.~\ref{images_red}.\\

\begin{figure*}
\vspace{-2.5cm}
\includegraphics[angle=-90,width=1.\linewidth]{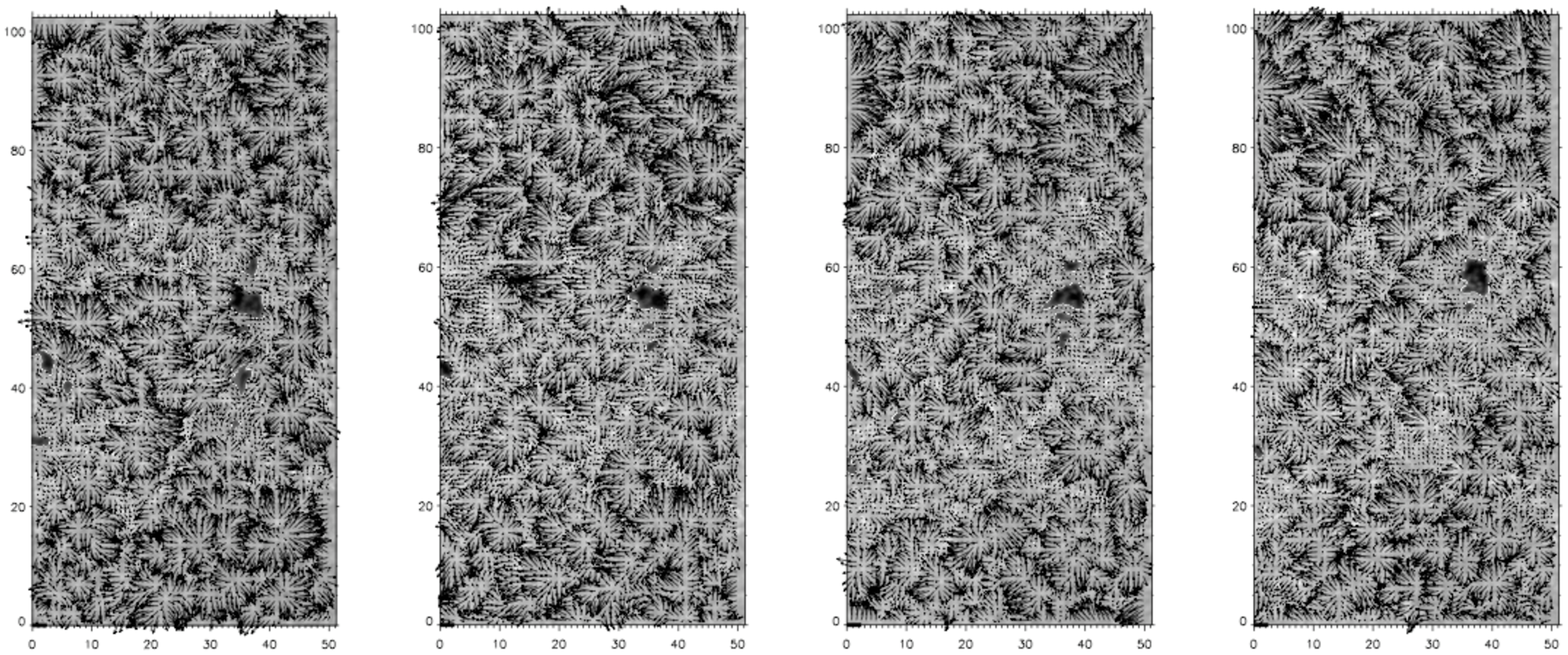}
\vspace{-2.5cm}
\caption{Map of horizontal velocities  (FWHM 1$\farcs$0) in the region displaying solar pores observed with \emph{Hinode}
on 30 Sep 2007. Fourteen maps are produced, each one computed from data in
1-hour intervals. From left to right, it is shown the maps corresponding to
intervals 1, 5, 9 and 13, respectively. The length of the black bar at
coordinates (0,0) corresponds to 2.3 km s$^{-1}$. The coordinates are expressed
in arc sec. The background represents the average image in every hour-set.}
\label{porosflowmapHINODEall}
\end{figure*}

Concerning \emph{Hinode} data, the computed map of horizontal velocities for
the time series of 1 June 2007 is displayed in Fig.~\ref{porosflowmapHinode}.
This map is calculated from the longest time series (67 minutes, see
Table~\ref{poroseries});  the analysis of the shortest series (not presented here) gives similar results though, as expected, noise and velocity magnitudes increase significantly. The isolated solar pore is immersed in a granular
region displaying several and recurrent large exploding granular events all
around it.  This velocity map as well as those derived from the SST data lead us to conclude  
that an imaginary line connecting the centers of the exploding events (i.e. mesogranules) around a pore 
outlines a round-shaped contour at a distance from the pore border comparable
to the pore diameter. Velocity magnitudes in the outer part of the outlined
contour are clearly larger than those in the inner part between the contour and
the pore border. The resulting map of horizontal velocities for 30 September
2008 data will be presented in Sect.~\ref{evol_map} where we pursue a detailed
analysis for this time series.\\

\subsection{Averaging horizontal flows within different temporal windows}
\label{averages}

To study the influence of different time averages in the velocity maps we
consider temporal windows of 5, 10, 15 and 20 min in the best quality G-band
time series recorded at the SST (series 2 of 30 September in
Table~\ref{poroseries}).  As expected, the maps are
smoother for longer averaging periods. The two extreme cases are shown in Fig.~\ref{LCTaver} displaying a close-up at some of the pores in the left part of the FOV in Fig.~\ref{porosflowmapSST} (\emph{lower panel}). Figure~\ref{histogram_averages} shows the histograms of the velocity magnitudes
for each averaging temporal window considered. The resulting distribution is
very similar in the four cases but the histograms shift to the left as 
the averaging period increases. Thus, the largest velocity magnitudes range
from 1.83 to 1.97 km s$^{-1}$ and the mean values from 0.48 to 0.50 km
s$^{-1}$. In both cases, smaller/larger values correspond to longer/shorter
time averages.

Regardless of whether we average over 5 or 20 min intervals, we find essentially the same general trends with exploding granular patterns all over the FOV, meaning that the averaging periods we are employing are smaller than or about the lifetime of the observed structures. We do not recognize any moat-like flow around the pores for such different time
averages as shown in Figure.~\ref{LCTaver}.

\subsection{Long-term evolution of the velocity field}
\label{evol_map}
In order to investigate how the evolution of the emerging region affects the
velocity in the FOV for long periods of time, particularly around solar pores,
we have used the \emph{Hinode} time series of 30 September 2007, which corresponds to
18 hours of almost continuous solar observation as reported in Sect.~\ref{30sep}. 
The stable good quality and long duration of the series is
ideal to study the evolution of the flow maps in an emerging flux region. To
that aim the images are grouped in 1-hour sets from  00:14 UT up to the time
14:00 UT (excepting the first set spanning from 00:14 UT to 00:59 UT),
and every set is processed independently resulting in 14 maps of
horizontal velocities which are computed by LCT using a Gaussian tracking window of
FWHM 1$\farcs$0 and 1-hour averages. Fig.~\ref{porosflowmapHINODEall} shows, out of these
fourteen maps, those starting at 00:14 UT, 05:00 UT, 09:00 UT and 13:00 UT, respectively. A movie displaying the evolution of the pore throughout
the 18-hour observation can be downloaded from the website
 \emph{http://www.iac.es/proyecto/solarhr/hinode30sep2007.mov}.\\

The main pore is located in the spatial position [37$\arcsec$,55$\arcsec$] in the first map
(Fig.~\ref{porosflowmapHINODEall}) and surrounded by some smaller pores which
form altogether a sort of vertical and elongated arrangement in the figure. The
collection of these pores is evolving in time and some of them start merging
and disappearing. The final picture of the region displayed in
Fig.~\ref{porosflowmapHINODEall} shows the isolated main pore with only a
very tiny magnetic companion.\\

We do not identify any signal of moat-like flow around the pores in any of the
evolutionary stages shown in the maps sequence but continuous activity caused
by exploding granules. Centers of divergence are systematically identified,
some of them very close to the pore border. Proper motions displaying inward
components are more common around the pores and {\bf no} outward regular 
large-scale flow, as corresponding to a moat-flow, is found. These results, the reliability 
of which is supported by the long duration of the sample studied, reinforce the 
previous analysis pursued for the only 1-hour long SST data in Sect.~\ref{horizontalpattern}.\\

\subsection{Distribution of horizontal speeds in the FOV}

\begin{figure}
\centering
\vspace{-1mm}
\begin{tabular}{c}
\vspace{-4mm}\hspace{-3.2cm}\includegraphics[width=.85\linewidth]{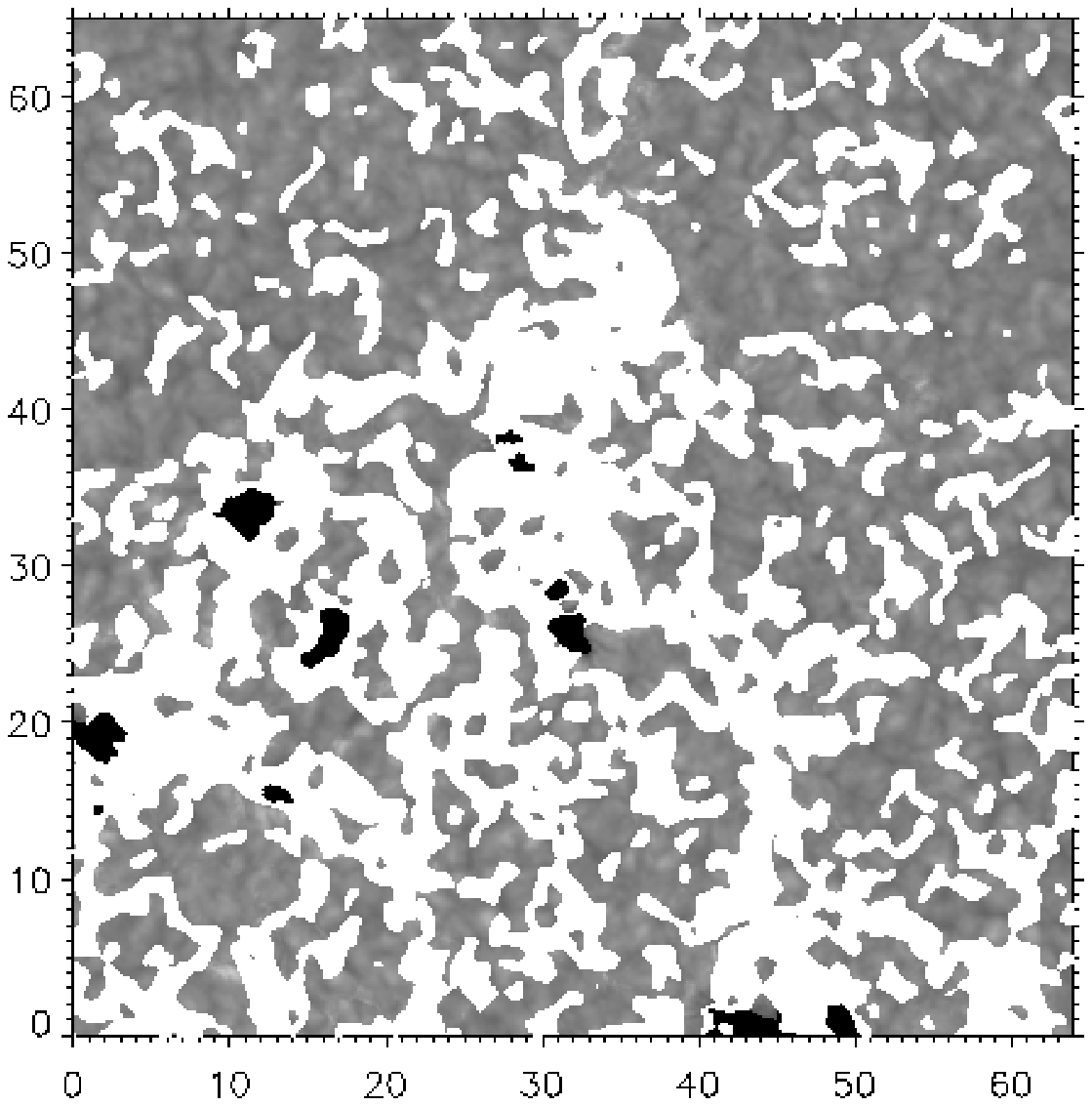} \\ 
\vspace{-4mm}\hspace{-3.2cm}\includegraphics[width=.85\linewidth]{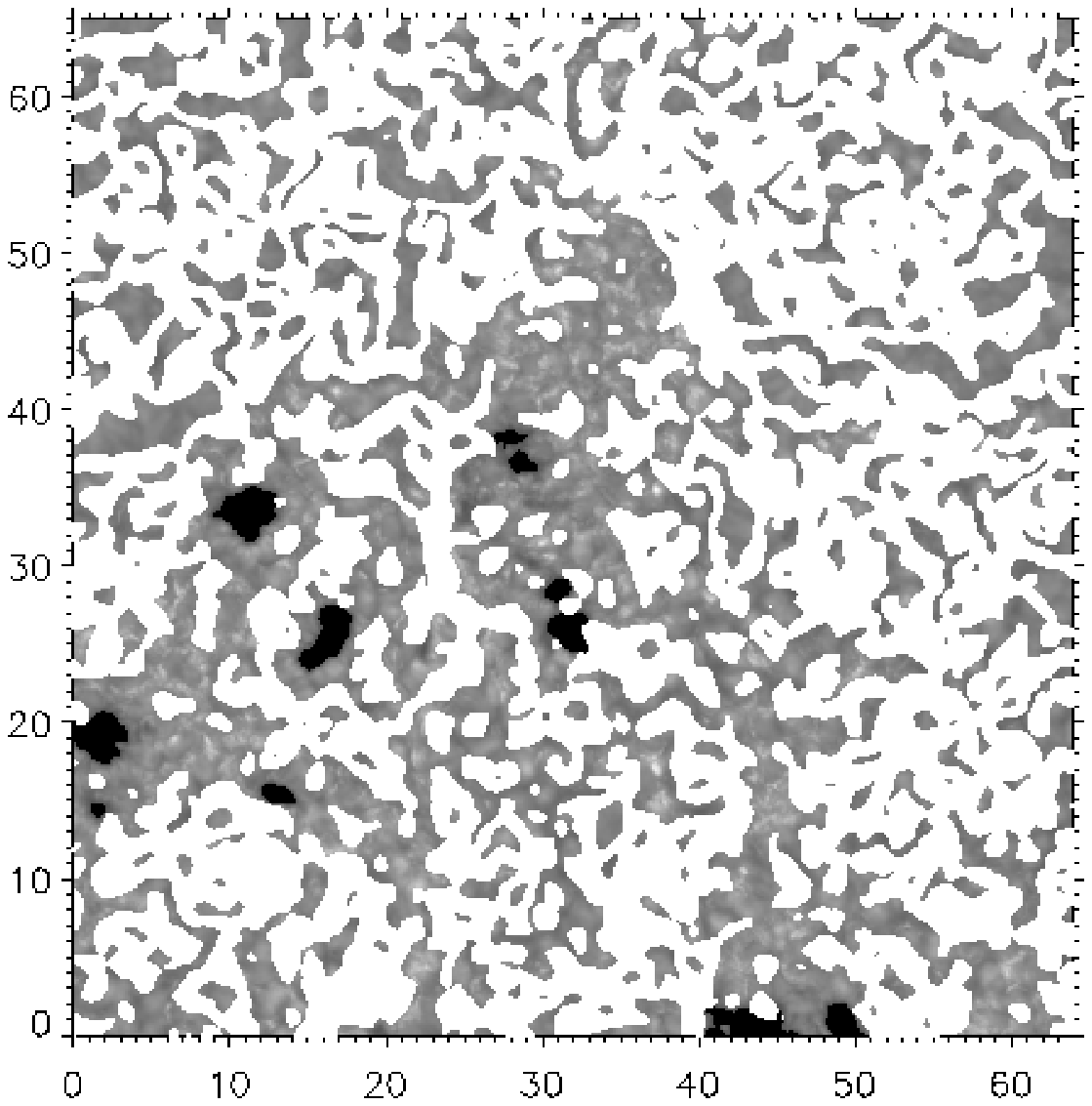} \\
\hspace{-1.5cm}\includegraphics[angle=90,width=1.3\linewidth]{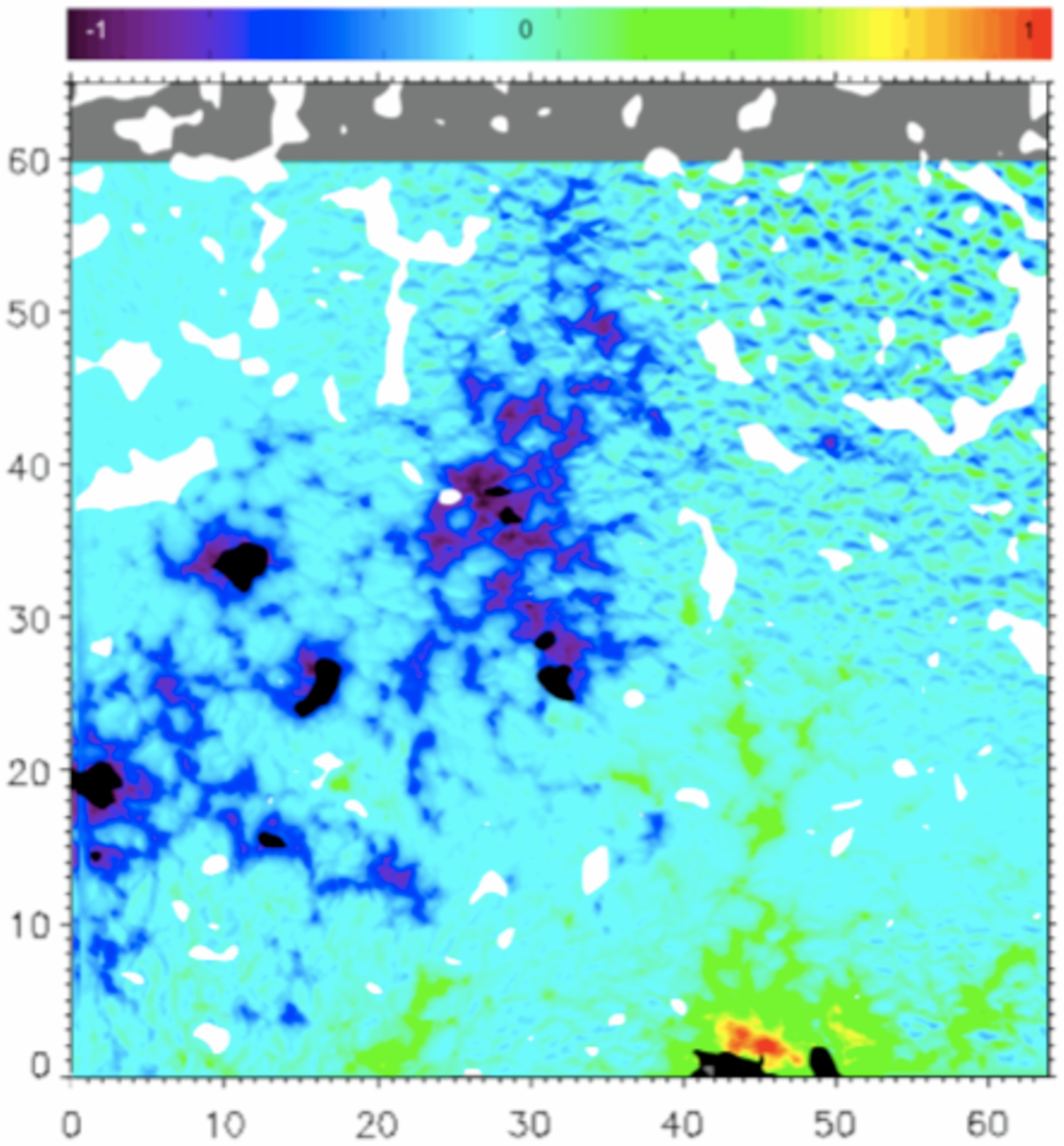}
\end{tabular}
\caption{Location of the areas (\emph{in white}) where the magnitude of
horizontal velocities is lower than 0.3 km s$^{-1}$ (\emph{upper panel}),
between 0.3 - 0.8 km s$^{-1}$(\emph{middle panel}) and greater than 0.8 km
s$^{-1}$(\emph{lower panel}).  Background images in the top and middle panels
correspond to a 48-min average image (SST G-band series 1, see Table~\ref{poroseries}), 
and in the bottom panel to a false-color normalized 
(factor of 2300 G) magnetogram of the region. Pores are colored in black. 
The coordinates are expressed in arc seconds.}
\label{vranges}
\end{figure}

Figure~\ref{vranges} shows the location (\emph{white areas}) of speeds
(velocity magnitudes) within three different ranges in km s$^{-1}$, for the SST
data (the first series in Table~\ref{poroseries}): low velocity magnitudes
lower than 0.3, medium velocity magnitudes in the range 0.3~-~0.8 and large
velocity magnitudes greater than 0.8.\\

Small speeds are mainly grouped in the central part of the FOV where an intense
magnetic activity is detected as evidenced by the high concentration of G-band
bright points and faculae present in this region (see Fig.~\ref{images_red}). 
Around pores, the velocity magnitudes mainly correspond
to the lower range \mbox{($<$ 0.3 km s$^{-1}$)} so that they are surrounded by
white areas in Fig.~\ref{vranges} (\emph{upper panel}). The areas mapping
medium velocity magnitudes are regularly spread out all over the FOV except in
the proximity of pores as shown in Fig.~\ref{vranges} (\emph{middle panel}).
Large velocities ($>$ 0.8 km s$^{-1}$) are not homogeneously distributed
in the FOV but mainly grouped at the two upper corners of Fig.~\ref{vranges} (\emph{lower panel}). 
The magnetogram in the last panel of Fig.~\ref{images_red} has 
been overlapped to better identify the match between the locations displaying larger
velocity magnitudes and the less-magnetized regions (scaled as 0 and colored in light blue). These large flows zones might
reflect the presence of supergranular cells as commented in Sect.~\ref{horizontalpattern}.\\

\subsection{Velocity distribution around solar pores}
\label{velpores}

\begin{figure}
\centering
\includegraphics[width=1.1\linewidth]{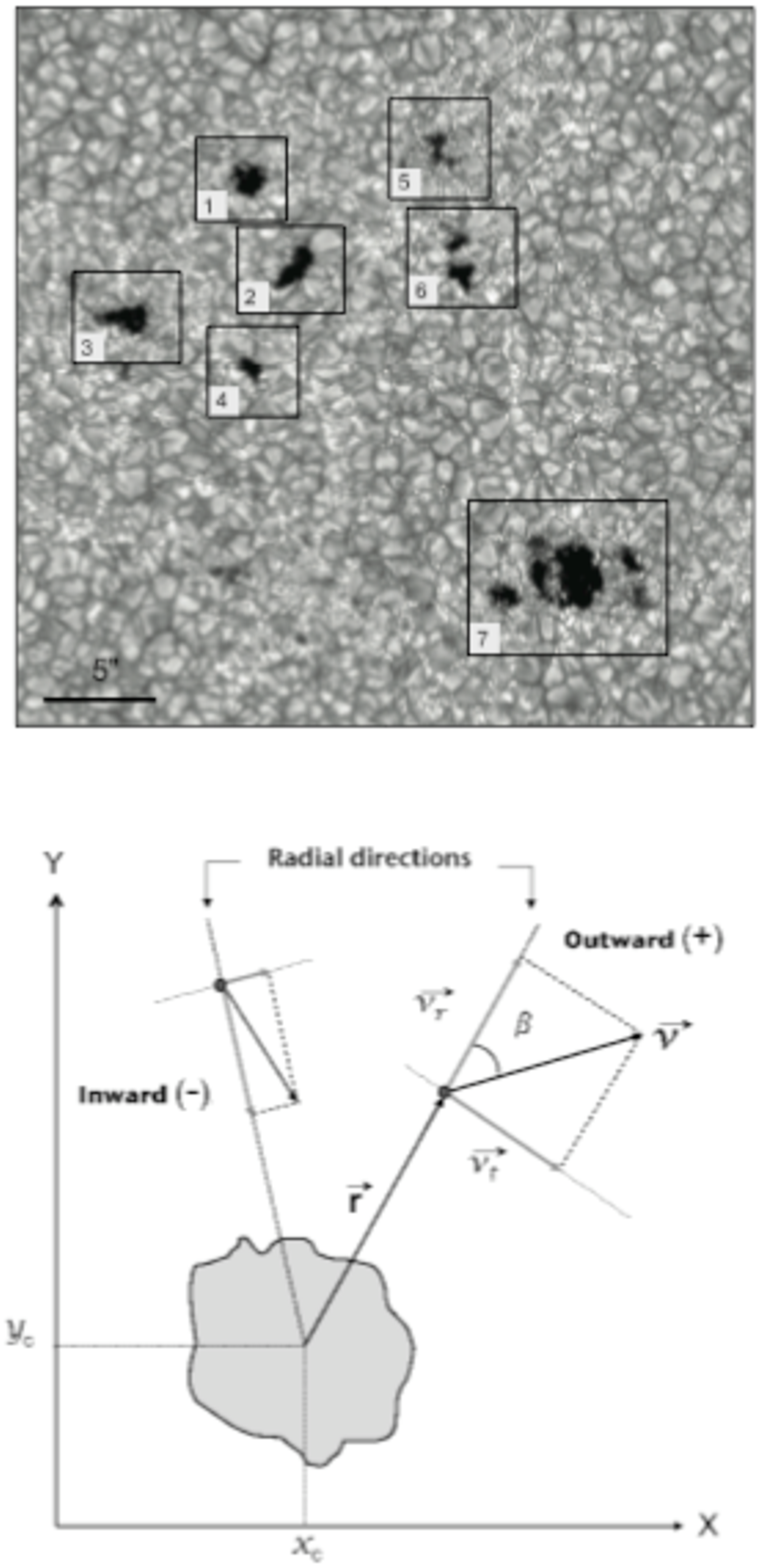}
\vspace{0cm}
\caption{\emph{Upper panel}: emerging active region observed with the SST (G-band) on 30
September 2007. \emph{Lower
panel}: sketch showing the projection applied to the velocities around a solar
pore centered at ($x_c,y_c$) respect to the orthogonal coordinate system  X, Y.
The figure shows the velocity vectors ${\bf v}$ for two points in the
granulation region around the pore (\emph{small black dots}).  The projection
of {\bf v} along the radial and transversal directions renders the radial $v_r$
and transversal $v_t$ velocity components, respectively.}
\label{canarypores}
\end{figure}
From the computed velocity fields we can perform a detailed analysis of the
velocity distribution around the solar pores. To pursue this study we will use
the region observed with the SST where we have a useful collection of pores available 
to work with. As we have two time series for this region (see Table~\ref{poroseries}),
we will employ the first one covering the longest time period, except for one of
the pores which is out of the FOV. For this pore, we will use the second time
series. Figure~\ref{canarypores} shows the FOV including all the pores under
study which are labeled with consecutive numbers for easy identification
hereafter.\\

\begin{figure*}
\vspace{-2cm}
\hspace{-11cm}\includegraphics[angle=-90,width=2.2\linewidth]{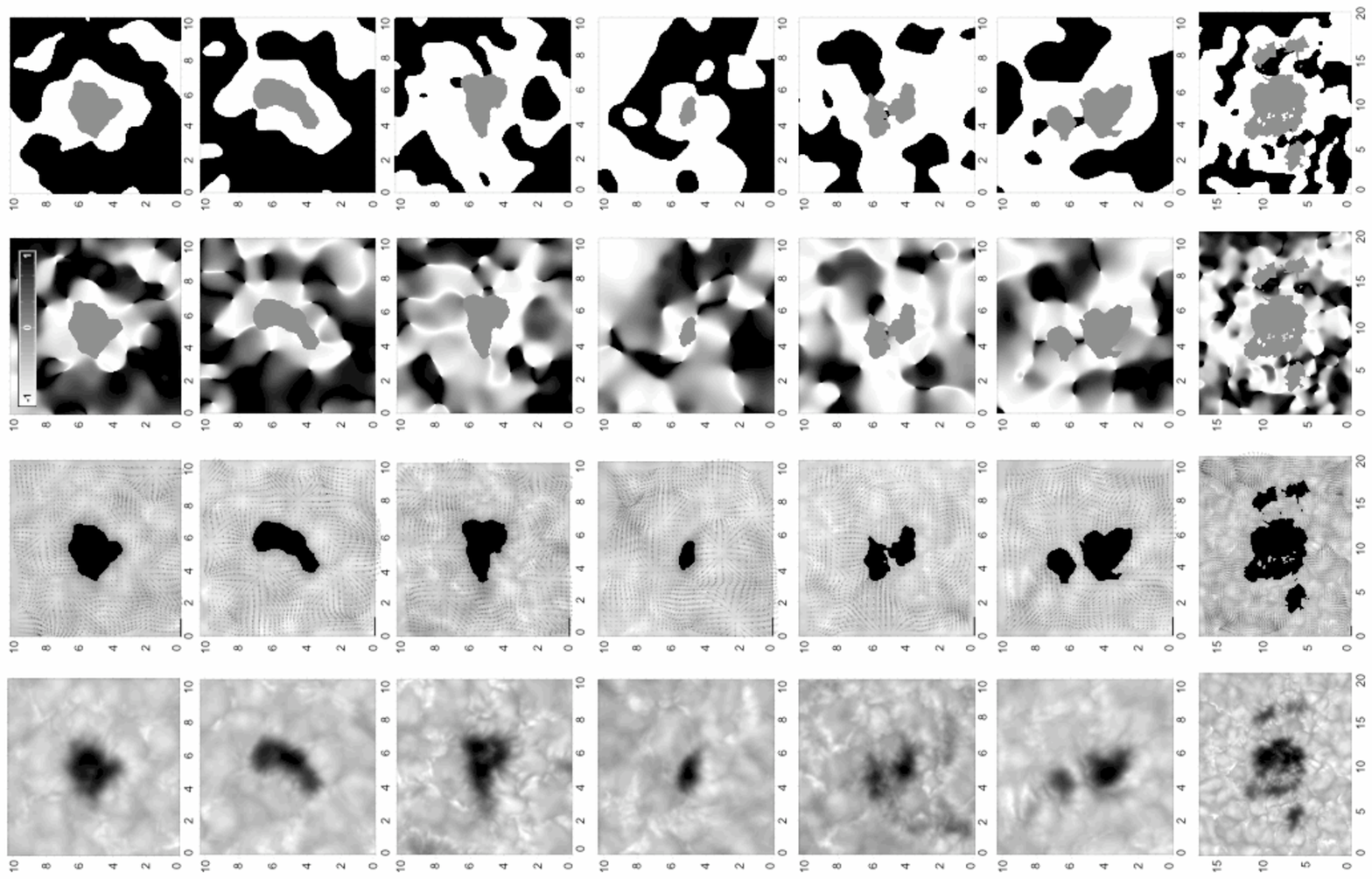}
\vspace{-2cm}
\caption{Discrimination between inward and outward motions surrounding solar
pores displayed in Fig.~\ref{canarypores} (\emph{upper panel}). Rows display
pores from 1 to 7 (\emph{top to bottom}) and the four columns correspond to
(\emph{left to right}): the average image of the time series, the map of
horizontal velocities, the map of $\cos\beta$ and the binary map of inward
(\emph{white}) and outward (\emph{black}) radial velocities. The length of the
black bar at coordinates (0,0) in the map of horizontal velocities corresponds to 1.6 km s$^{-1}$. The gray bar in the top $\cos\beta$ map 
shows the scale for the cosine from -1 ($\beta=180^\circ$) to 1 ($\beta=0^\circ$). See Fig.~\ref{canarypores} (\emph{lower panel}) and the text for details. The spatial units are in arc sec.}
\label{radialvel_pore}
\end{figure*}

Figure~\ref{canarypores} (\emph{lower panel}) illustrates the projection of the
velocities into radial and transversal components as a convenient way to
compute inward and outward motions. The figure plots two points in the
granulation surrounding a solar pore with their corresponding velocity vectors
${\bf v}$. The pore is centered at coordinates $(x_c,y_c)$ with respect to the
orthogonal coordinate system X,Y placed at the lower left corner of the FOV.
Vector ${\bf r}$ is the position vector of a given point with respect to the
pore center. The pore center is located at its gravity-center calculated by
weighting the position of every point inside the pore with the inverse of its
respective intensity. Velocity vectors in every point of the granulation
surrounding the pore are projected into radial $v_r$ and transversal $v_t$
components.\\

In order to establish the inward and outward motions, we first select the FOV
including the pore under study. Since active regions in general, and pores in
particular, exhibit their own displacement (due to differential rotation and
intrinsic motions) while embedded in the granulation pattern, we align the time
series with respect to an area framing the pore (correlation box) so that we
make sure we are measuring plasma motions with respect to the pore. We compute
the map of horizontal velocities by LCT with the same Gaussian tracking window
(FWHM 1$\farcs$0) employed in the previous sections. The next step is to define
the "radial directions" (see lower panel in Fig.~\ref{canarypores}) that will
be used as the reference to project the velocities. For short distances these
directions are defined as perpendicular to the pore border. These perpendicular
directions are calculated from the gradients of intensity in a smoothed
pore-mask image\footnote{A binary mask setting the area occupied by the pore
(1/0 inside/outside the pore) is defined. Then this mask is smoothed by
convolving with a Gaussian function so that we obtain a distribution of
pseudo-intensities smoothly ranging from 1 to 0 and describing a blurred shape
of the pore. This will be the smoothed pore-mask image to which we refer
above.}. This way, one can also deal with non round-shaped pores. The limit for
short distances is defined by thresholding the intensity gradients. The
threshold depends on the pore shape and size. At large distances 
(i.e.~$\sim$3 times larger than the pore mean radius) all pores are considered as
round-shaped structures and the radial directions
are defined by the position vector ${\bf r}$ of a given point with respect to
the pore center.  According to the lower panel in Fig.~\ref{canarypores},
inward/outward motions correspond to $\cos\beta$ negative/positive, where
$\beta$ is the angle formed by $\bf v$ and the positive radial direction
(outward) at each point of the FOV. The value of $\cos\beta$ is mapped in gray
scale ranging from 1 for purely radial outward velocities (\emph{in black}) to
-1 for purely inward velocities (\emph{in white}). A binary mask is created
from the previous gray-scaled map where areas in \emph{black} and \emph{white}
correspond to velocities with positive (outward) and negative (inward) radial
components, respectively.\\

\begin{figure}
\centering
\vspace{-2cm}
\includegraphics[width=1.\linewidth]{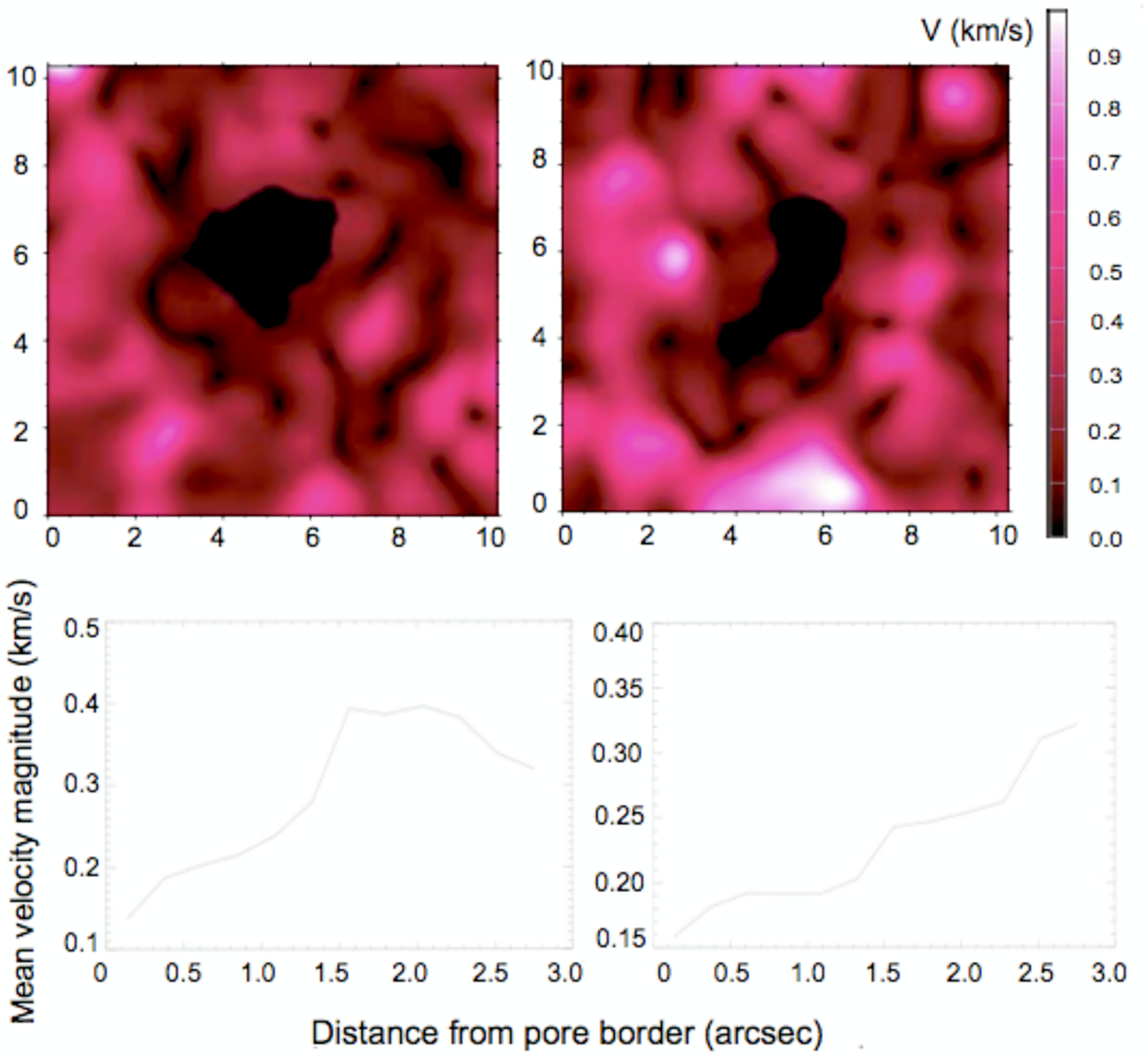}
\vspace{-2cm}
\caption{Analysis of the velocity magnitudes of proper motions around the first to \emph{regular-shaped} solar
pores in Fig.~\ref{radialvel_pore}. \emph{Upper panels}:  magnitude of horizontal velocities in false-color
representation. \emph{Lower panels}: plot of the mean velocity magnitude versus
the distance from the pore border. The spatial units are in arc sec.}
\label{magvel_pore}
\end{figure}

We apply the aforementioned method to our sample of 7 different pores shown in
the upper panel of Fig.~\ref{canarypores}. The results are displayed in
Fig.~\ref{radialvel_pore} where rows display the different pores and the
columns correspond to: 1) the averaged FOV around the pore; 2) the velocity
field; 3) the gray-scale representation of $\cos\beta$; and 4) the final binary
representation mapping the areas with inward (\emph{white}) and outward
(\emph{black}) radial velocity components, after applying the binary mask. The
analysis of all cases establishes that the flows display a clear preference for
inward directions around the pores. This fact is systematically found in all
examples. The more regular-shaped pores are surrounded by an also regular
annular-like area with inward velocity components, having a mean width similar
to the center-to-border distance in the pore. The dividing line between inward
and outward motions is connecting the centers of divergence.\\
Fig.~\ref{magvel_pore} (upper panels) shows color-scaled maps of the velocity
magnitudes (ranging from 0 to 1 km s$^{-1}$) for the more regular-shaped pores
(1 and 2) in our sample, represented in the the two upper rows of
Fig.~\ref{radialvel_pore}. Note that both pores exhibit proto-penumbral
structures which are also visible in other pores of the sample. The centers of
divergence are clearly identified in the maps of Fig.~\ref{magvel_pore} as
\emph{black structures}  around the pores. Another important distinctive
feature in these maps is that the highest speeds are located beyond these black
structures away from the pore.\\
 
In an attempt to study a possible systematic trend in the variation of the
speeds around the most regular-shaped pores having approximately the same size  in our sample, as a first approach and for simplicity, we calculate the mean velocity magnitude in consecutive
ribbons (strips) encircling the pores 1 and 2. A total of 12 adjacent ribbons
($\sim$ 0$\farcs$24 wide), referred to as ring-like structures surrounding the
pore, are used for this analysis. 

We emphasize that the following results should not be taken as trying to globalize the properties
and behavior of the flows around all solar pores but as a preliminar and simplistic case in which we can compare
two pores of our sample under similar conditions (size, shape, polarity among others) and apply the
above-mentioned method of computing velocities in ribbons around them. In a future work we will conduct 
a study to be able to compare more general cases.

Figure~\ref{magvel_pore} (\emph{lower panels}) plots the speeds versus
distance to the pore border. The velocity magnitudes increment as a function of
this distance: 1) the mean value of speeds increments rapidly at distances
ranging from $\sim$ 0$\farcs$15 to 0$\farcs$4; 2) in the range
0$\farcs$4~-~1$\farcs$0 the variation curve is flatter; 3) in the range
1$\farcs$0~-~1$\farcs$6 we again obtain very sharp increments. Up to this
distance from the pore border, we observe the same behavior in both pores.
Nevertheless, the mean speed values at 1$\farcs$6 are quite different (250 vs.\
390 m s$^{-1}$). At further distances ($>$ 1$\farcs$6) the trend in the mean
velocity magnitudes differs substantially. We must bear in mind that the flows
around the pore can also be affected by the intrinsic characteristics of every
single pore and by the contribution from other sources in the neighbourhood,
e.g.\ pores in the vicinity.

\section{Conclusions}
\label{sec:dis}
The proper motions in solar active regions displaying pores are analyzed from
high-resolution time series of images. The observing material stems from
coordinated ground-based and space observations. Thus, part of this material
was acquired with the Swedish 1-m Solar Telescope, and reconstructed by
employing the novel MFBD and MOMFBD techniques to achieve image resolutions
near the diffraction limit. The other part of the data stems from the solar
telescope on board the \emph{Hinode} satellite. The long duration, stability
and high-resolution of the time series achieved by \emph{Hinode} enable us to
study dynamical properties of the photospheric horizontal flows along periods
of time much longer than those typically reachable from ground-based observations
which are restricted by varying seeing conditions.\\

The local correlation technique applied to the time series allowed us to track
the proper motions of structures in solar active regions and particularly in
the areas nearby solar pores. Proper motions have been tracked in a variety of
active regions for periods of typically 20-60 min but also one
for several hours. We conclude that the flow patterns derived from different
observational sets are consistent among each other in the sense that they show
the determinant and overall influence of exploding events in the granulation
around the pores and in the whole FOV. Motions toward the pores in their
nearest vicinity are the dominant characteristic we claim to observe
systematically. Thus, we do not find any trace of moat flow in the wide sample
of pores studied. The motions at the periphery of the pores are basically
influenced by the external plasma flows deposited by the exploding events, as
suggested by other authors in previous works  \citep{sobotka1999, roudier2002, sankara2003}.
In addition, the horizontal velocity magnitudes are clearly lower ($<$ 0.3 km s$^{-1}$) 
in the nearest locations surrounding the pores and, in general, in the more magnetized regions
in the FOV, as expected due to the inhibition of convection taking place.\\

Our results are also in agreement with recently developed 3D radiative magnetohydrodynamic simulations of pore-like magnetic structures that report downflows surrounding them, maintained by 
horizontal flows towards the simulated pore \citep{cameron2007}. Moreover, we interpret the dividing line between radial inward and outward motions, found
by \cite{deng2007} outside the residual pore in the last stage of a decaying
sunspot, as corresponding to the location of the centres of divergence of the
exploding events around the pore. The outward motions these authors describe,
which are not in the immediate surroundings of the pore but separated by the
annular inward motion, would then correspond not to moat flows but to the
outward flows originated in the regular mesh of divergence centers around the
pore. \\

\begin{acknowledgements}
The Swedish 1-m Solar Telescope is operated in the island of La Palma by the Institute of Solar Physics of the Royal Swedish Academy of Sciences in the Spanish Observatorio del Roque de los Muchachos of the Instituto de Astrof\'isica de Canarias. We thank the scientists of the \emph{Hinode} team for the operation of the instruments. \emph{Hinode} is a Japanese mission developed and launched by ISAS/JAXA, with NAOJ as domestic partner and NASA and STFC (UK) as international partners. It is operated by these agencies in co-operation with ESA and NSC (Norway). The reconstruction of images using the MOMFBD technique is a very computationally expensive task. To reduce the time needed for it, we made use of the Condor workload management system (http://www.cs.wisc.edu/condor/). S. Vargas thanks the people from the Institute of Solar Physics in Stockholm for his support during his short-stay in Sweden. Partial support by the Spanish Ministerio de Ciencia e Innovaci\'on through projects ESP2006-13030-C06-01 and AYA2007-63881, and financial support by the European Commission through the SOLAIRE Network (MTRN-CT-2006-035484) are gratefuly 
\end{acknowledgements}

\end{document}